\documentclass[aps,pre,groupedaddress,amsmath,amssymb]{revtex4-1}
\usepackage{graphicx}  % needed for figures
\usepackage{dcolumn}   % needed for some tables
\usepackage{bm}        % for math
\usepackage{verbatim}   % for math
\usepackage{color}
\usepackage[utf8]{inputenc}

\begin{document}

\title{Locomotion of a flexible one-hinge swimmer in Stokes regime}
\author{Priyanka Choudhary}
\affiliation{
   Department of Physics,
   Malaviya National Institute of Technology Jaipur,
   Jaipur, 302017 (India) \\
   Department of Physics,
   Indian Institute of Technology Delhi,
   Hauz Khas, New Delhi 110016 (India)
   }
   
   \author{Subhayan Mandal}
\affiliation{
   Department of Physics,
   Malaviya National Institute of Technology Jaipur,
   Jaipur, 302017 (India)
   }

\author{Sujin B. Babu}
\affiliation{
   Department of Physics,
   Indian Institute of Technology Delhi,
   Hauz Khas, New Delhi 110016 (India)
   }

\date{\today}
\begin{abstract}
E. M. Purcell showed that a body has to perform non-reciprocal motion in order to propel itself in a highly viscous environment. The swimmer with one degree of freedom is bound to do reciprocal motion, whereby the center of mass of the swimmer will not be able to propel itself due to the Scallop theorem. In the present study, we are proposing a new artificial swimmer called the one hinge swimmer. Here we will show that flexibility plays a crucial role in the breakdown of Scallop theorem in the case of one-hinge swimmer or two-dimensional scallop at low Reynolds number. To model a one-hinge artificial swimmer, we use bead spring model for two arms joined by a hinge with bending potential for the arms in order to make them semi-flexible. The fluid is simulated using a particle based mesoscopic simulation method called the multi-particle collision dynamics with Anderson thermostat. Here we show that when our swimmer has rigid arms, the center of mass of the swimmer is not able to propel itself as expected from the Scallop theorem. When we introduce flexibility in the arms, the time reversal symmetry breaks in the case of the one-hinged swimmer without the presence of a head contrary to the one-armed super paramagnetic swimmer which required a passive head in order to swim. The reduced velocity of the swimmer is studied using a range of parameters like flexibility, beating frequency and the amplitude of the beat, where we obtain similar scaling as that of the one-armed super paramagnetic swimmer. We also calculate the dimensionless Sperm number for the swimmer and we get the maximum velocity when the Sperm number is around $\sim 1.8$
 \end{abstract}

\maketitle

\section{Introduction}
Microorganisms have adapted their locomotion to the harsh environment of low Reynolds number regime by invoking different swimming strategy \cite{Eric_lauga_2009}. For example, the E. Coli moves moves by rotating its helical flagellum \cite{Ecoli_nature_1973,berg2005coli}, Chlamydomonas flagella have a breast stroke kind of motion \cite{Chlamydomona_2000}, African Trypanosome has a helical flagellum attached to the cell body with a planar wave passing through it \cite{Oberholzer_2010,Babu_trypanosome} etc. Swimming of these kind of natural swimmers have been investigated for the last half-century \cite{Eric_lauga_2009,Ecoli_nature_1973,bray2001cell,berg2005coli,Chlamydomona_2000,Oberholzer_2010,Babu_trypanosome,gray1955propulsion,pacey1994acquisition}. As a result of these studies, artificial swimmers have also been proposed, like Taylor sheet \cite{Taylor_1951}, Purcell's two-hinge swimmer \cite{Purcell_1977,Avron_2008}, three-linked spheres swimmer \cite{Ali_2008}, Elastic two-sphere swimmer \cite{PhysRevFluids.2.043101} and Three-sphere with a passive elastic arm \cite{Montino2015} etc, which have further enhanced our understanding about low Reynolds number swimmers. One of the challenges in proposing an artificial swimmer lies in the fact that the proposed movement stroke should not be reciprocal otherwise it cannot propel itself due to the Scallop theorem. In Scallop theorem, Purcell had argued that a swimmer with one-hinge or one degree of freedom is bound to perform reciprocal motion and thus will not be able to swim in the Stokes regime \cite{Eric_lauga_2009, Purcell_1977}.
\subparagraph{}
Purcell proposed two possible ways to elude from Scallop theorem, one is "corkscrew" \cite{Ecoli_nature_1973,purcell1997efficiency} motion and second is "flexible oar" \cite{Goldstien_1998,lagomarsino2003simulation} motion. Using the concept of flexible oar, Dreyfus et al. \cite{magnetic_swimmeri_nature_2005} reported a micro swimmer that exploit elastic property of a slender filament made up of paramagnetic beads. To break the time inversion symmetry, a passive head was attached to the flexible arm. The passive head reduces the velocity of the flexible swimmer, bigger the head, higher is the drag force experienced by the swimmer. The head is essential for swimming because without it the tail performs a reciprocal motion and the  velocity of the swimmer reduces to zero \cite{lauga2007floppy}.
\subparagraph{}
     In the present study, we design and simulate a two-dimensional swimmer having two symmetric arms joined by a single hinge that looks like a two-dimensional scallop. The arms of the swimmer are semi-flexible and thus behave as an elastic scallop. To design an elastic scallop swimmer we use bead spring model \cite{babu2011dynamics} for the two arms joined by a common bead which acts as a hinge and to introduce flexibility to the two symmetric arms of the swimmer, we use bending potential \cite{allen-tildesley-87}. The actuation happens only at the hinge and the rest of the arm relaxes depending on the strength of the actuation similar to the one-armed flexible swimmer \cite{Goldstien_1998}. In case of the elastic scallop the time inversion symmetry is broken because through the viscous drag term time enters into the equation of the filament shape, similar to the one-armed swimmer. Hydrodynamic interactions between the arms of the swimmer are implemented using the particles based simulation method for the fluid called as multi-particle collision dynamics (MPC) \cite{Gompper2009}. The advantage of MPC is that it solves the Navier-Stokes equation and also has inherent thermal fluctuations \cite{malevanets1999mesoscopic,malevanets2000solute} resembling real fluids. By this method a variety of hydrodynamic problems have been solved, for example, swimming of sperm cells \cite{Sperm_mpc_2008}, African Trypanosome \cite{Babu_trypanosome}, E. coli \cite{hu2015modelling}, Taylor line \cite{munch2016taylor}, Spheroidal microswimmer \cite{theers2016modeling}, Squirmer \cite{downton2009simulation,zottl2012nonlinear} etc. This is the method of our choice as it is very easy to implement and is also shown to be one of the most efficient method in the Stokes limit \cite{Gompper2009}.
\subparagraph{}
 We know that the dynamics of the one-armed swimmer is described by the non dimensional hyper diffusion equation given by Wiggins et al. \cite{wiggins1998trapping,Goldstien_1998}. They have already shown that a flexible slender filament is propelled by periodic actuation, which is characterized by a dimensionless number called the Sperm number $S_p$ \cite{Goldstien_1998}. Sperm number is the ratio of the length of the swimmer to its hydrodynamic penetration length. When $S_p\ll 1$, the penetration length is larger than the length of the swimmer, which means that the arms of the swimmer are stiff and perform near reciprocal motion, because of that we get very small velocity. When $S_p\gg 1$, we have the length of the swimmer to be very large compared to the penetration length, and hence the drag forces acting on the swimmer increase, thus the velocity of the swimmer again reduces. The interplay between these two effects leads to a maximum in the velocity at a point when the hydrodynamics penetration length is approximately equal to the length of the swimmer as shown by Wiggins et al. \cite{wiggins1998trapping,Goldstien_1998}. 
 \subparagraph{}
      The paper is arranged as follows. In section \ref{sec:2} we describe the model of our two-dimensional swimmer using bead spring and bending potentials. Then we briefly describe the particle based simulation technique MPC and also explain how we couple the swimmer with the solvent particles to properly resolve hydrodynamic interactions. In section \ref{sec:3} we discuss how our one-hinge swimmer breaks time inversion symmetry with flexible arms. We also study the velocity of the swimmer as a function of bending potential, frequency of actuation and amplitude of the actuating wave. We are able to define the dimensionless Sperm number for this swimmer and also discuss how our dimensionless velocity vary with respect to $S_p$. In section \ref{sec:4} we discuss how this artificial swimmer can be realized experimentally, followed by the conclusions in section \ref{sec:5}.
\section{COMPUTATIONAL METHODS}
\label{sec:2}
\subsection{Multiparticle collision dynamics (MPC)}
To simulate our artificial swimmer in two-dimensional Newtonian fluid, we make use of the coarse-grained simulation technique called multi-particle collision dynamics (MPC). Although there are different variation of MPC, we use the technique MPC with Anderson thermostat, where both linear and angular momentum are conserved \cite{Gompper2009}. We know that MPC solves the Navier-Stokes equation and has been shown to be computationally efficient method for low Reynolds number hydrodynamics \cite{Gompper2009}. MPC is a particle based method where we use fictitious point particles of mass $m_0$ to mimic the fluid. MPC consists of two steps, one is called the streaming step and another one is called the collision step. In streaming step, we update the position $\textbf{r}_i(t)$ of all the solvent particles at time $t$ according to
\begin{equation}
 \textbf{r}_i(t+\delta t)= \textbf{r}_i(t) + \delta t \textbf{v}_i(t).
\label{e1}
\end{equation}
Where $\textbf{v}_i(t)$ is the velocity of $i_{th}$ particle at time $t$ and $\delta t$ is the MPC time step.
\subparagraph{}

The collision step is performed by dividing the simulation box into the square cells of length $a_0$. In the present study, we keep the number of fluid particles per cell $\rho=10$ with periodic boundary condition. The initial velocities of the particles are assigned from a Gaussian distribution with variance $k_BT/m_0=1$. In collision step, the velocity of each particle $\textbf{v}_i(t)$ is updated according to  
\begin{equation}
\begin{split}
\textbf{v}_i(t&+\delta t) =\textbf{u}(t) + \textbf{v}_i^{ran} - \sum_{i\in cell} \textbf{v}_i^{ran}/N_c + \\
&\quad \left\{m_0\textbf{I}^{-1} \sum_{j\in cell} \textbf{[r}_{j,c} \times (\textbf{v}_j - \textbf{v}_j^{ran})\textbf{]}\times \textbf{r}_{i,c}\right\}.
\label{e2}
\end{split}
\end{equation}
Where $\textbf{u}(t)$ is the center of mass velocity of all particles in a cell, $\textbf{v}_i^{ran}$ is a random velocity taken from a Gaussian distribution with variance $k_BT/m_0=1$, $N_c$ is the number of particles in a cell, $\textbf{I}$ is the moment of inertia of all the particles in a cell, $\textbf{r}_{i,c} = \textbf{r}_i - \textbf{R}_c$ is the relative position of $i_{th}$ particle in a cell with respect to the center of mass position $\textbf{R}_c$ of all the particles in a cell. 
\subparagraph{}

The swimmer keeps on actuating thereby pumping energy into the fluid which increases the kinetic energy of the fluid particles, and therefore the temperature of the fluid also increases known as viscous heating. The Anderson's thermostat which is inherent in the simulation technique, keeps the temperature of our system constant. In collision rule MPC conserves both linear as well as angular momentum. In MPC simulation, we know that when $\delta t \ll 1$, fluid particles become correlated thereby the system no longer remains Galilean invariant.  In order to restore Galilean invariance we perform a random shift of the simulation box in the interval $[-a_0/2,a_0/2]$ as proposed by Ihle et al. \cite{ihle2001stochastic,ihle2003stochastic}.  
\subparagraph{}
In present work, we have measured quantities in MPC units, where length is normalised with $a_0$, mass with $m_0$, energy with $k_BT$ and time with $\tau_0=a_0\sqrt{m_0/k_BT}$ \cite{padding2006hydrodynamic}. We have used $a_0=1$, $m_0=1$ and $k_BT=1$, which makes time unit equals to unity. One of the advantage of MPC is that, we have analytical expressions for the calculation of transport coefficients \cite{Gompper2009}. The total kinematic viscosity $\mu$ is the sum of the kinetic viscosity $\mu^{kin}$ and collision viscosity $\mu^{col}$, where $\mu^{kin} = 0.61 \delta t$ and $\mu^{col} = 0.036 / \delta t$. We have used $\delta t =0.01$, which keeps the kinematic viscosity $\mu\approx3.6$.   
           
\subsection{Modelling two-dimensional scallop}
\graphicspath {}
\begin{figure}
\centering
  \includegraphics[scale=0.18]{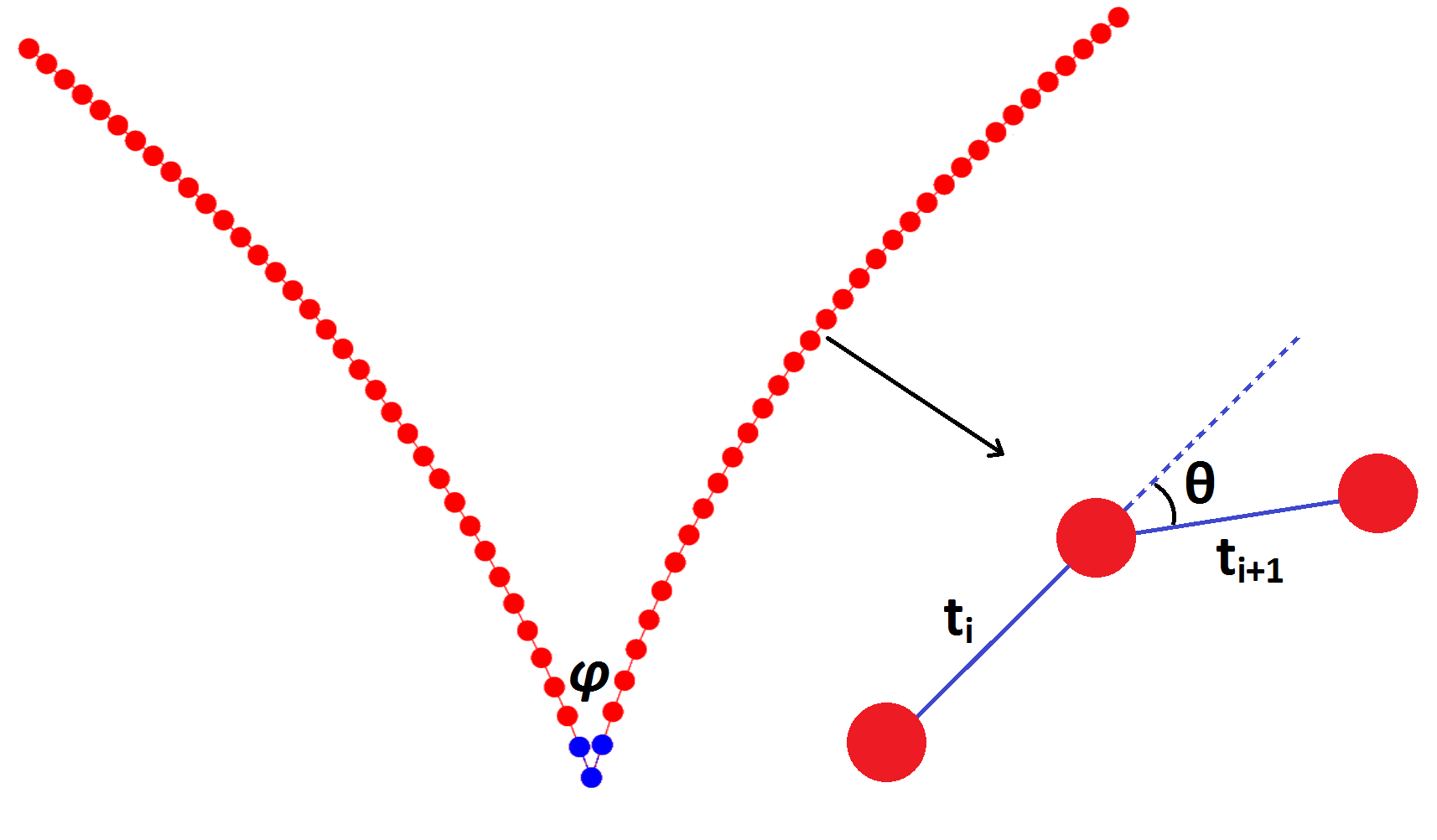}
\caption{The snapshot of a model one-hinge artificial swimmer with flexible arms. Three mass points are given blue color to show that the bending waves pass through only these three points. We have magnified the bond vectors $t_i$ and $t_{i+1}$, which connect neighboring mass points and make an angle $\theta$. $\varphi$ is the angle between two arms near the hinge.}
\label{fig1}
\end{figure} 

In order to model the two-dimensional swimmer with one-hinge, we consider a string of $N$ odd numbered mass points of mass $m_0=1$ connected with spring \cite{babu2011dynamics}. The middle bead acts as a hinge for the two-armed swimmer, so that the length of each arm is the same as shown is figure \ref{fig1}. The spring potential between the mass points is given by  
\begin{equation}
 U_s = \frac{1}{2} K_s (l-l_0)^2.
 \label{eq.3}
 \end{equation}
Where $K_s$ is the spring constant, $l$ is the distance between two mass points at any given time and $l_0$ is the equilibrium distance between the two mass points. In our simulation we keep the equilibrium spring length $l_0 = 0.5 a_0$. To make sure that the springs remain relatively stiff during the simulations, we use a high spring constant $K_s = 10^8$ in the present study. The arms of the swimmer are made semi-flexible in nature by introducing a three body bending potential
\begin{equation}
U_b = K_b (1-\cos \theta).
 \label{eq.4}
 \end{equation}
Where $K_b$ is the stiffness constant, $\theta$ is an angle between two bond vectors $\textbf{t}_i$ and $\textbf{t}_{i+1}$, as illustrated in figure \ref{fig1}(magnified bond vectors) \cite{babu2011dynamics}. We apply the bending potential along both the arms of the swimmer except for the hinge points. When $K_b \rightarrow \infty$, we have rigid arms for the swimmer and for $K_b \rightarrow 0$, we have completely flexible arms.  We vary the value of $K_b$ from $10^4$ to $10^7$, thereby vary the rigidity along the arms of the swimmer. 
\subparagraph{}

In order to simulate a two-dimensional scallop we define an angle $\varphi$ (see figure \ref{fig1}) between the two arms of the swimmer, which oscillates between a minimum and maximum angle. In order to achieve this, we introduce a three body bending wave potential along the three blue colored beads given by 
\begin{equation}
U_w = \frac{1}{2} K_w [t_{i+1}-\textbf{R}(l_0 \alpha)t_i]^2.
 \label{eq.5}
 \end{equation}
Where $K_w$ is a bending stiffness constant that decides the strength of the potential, $\textbf{R}$ is the rotation matrix, $\phi = l_0 \alpha $, $\alpha = A\sin^2 {(2\pi\nu t)}$ is the spontaneous curvature between the mass points where the bending wave potential is applied, $A$ is the amplitude of the wave and $\nu$ is the frequency with which the potential makes the beads to beat \cite{Babu_trypanosome}. The matrix $\textbf{R}$ rotates one bond vector $\textbf{t}_i$, against the neighboring bond vector  $\textbf{t}_{i+1}$,  about a unit vector which is perpendicular to $\textbf{t}_i$ and $\textbf{t}_{i+1}$, by an angle $\phi$. As the curvature $\alpha$ is function of square of the sine wave, we will have only positive values for curvature, thus the angle $\varphi$ between the arms can vary only between $0 - \pi$. When two arms are completely opened (i.e. zero curvature position), the angle $\varphi$ will be the maximum $\varphi_{max}= \pi$ at that time, the value of $K_w=4 \times 10^5$ is used in the present work. The total force on the $i_{th}$ bead of the swimmer, due to spring and bending potentials is given by $\textbf{F}_i = - {\nabla}_i(U_s+U_b+U_w)$. The movement of the swimmer is implemented by molecular dynamics (MD) technique. We update the velocities and positions of the beads using a leap frog velocity Verlet algorithm \cite{allen-tildesley-87}, where the integration time step is always kept at $\delta t_{MD}=10^{-4}$.

\subsection{Incorporation of swimmer with fluid particles}
  We perform $n$ number of MD steps, where $n=\delta t /\delta t_{MD}$ before a MPC step. In streaming step of MPC, we only consider the fluid particles, and during this step we allow the fluid particle to pass through the swimmer as well. We in-cooperate the mass points of the swimmer with the fluid particles in the collision step, which give the correct hydrodynamics for the swimmer \cite{Yeomans,noguchi2005dynamics,babu2011dynamics}. 
\section{RESULTS} 
\label{sec:3}

\subsection{Breakdown of the time inversion symmetry}
\begin{figure}
\centering
  \includegraphics[scale=0.32]{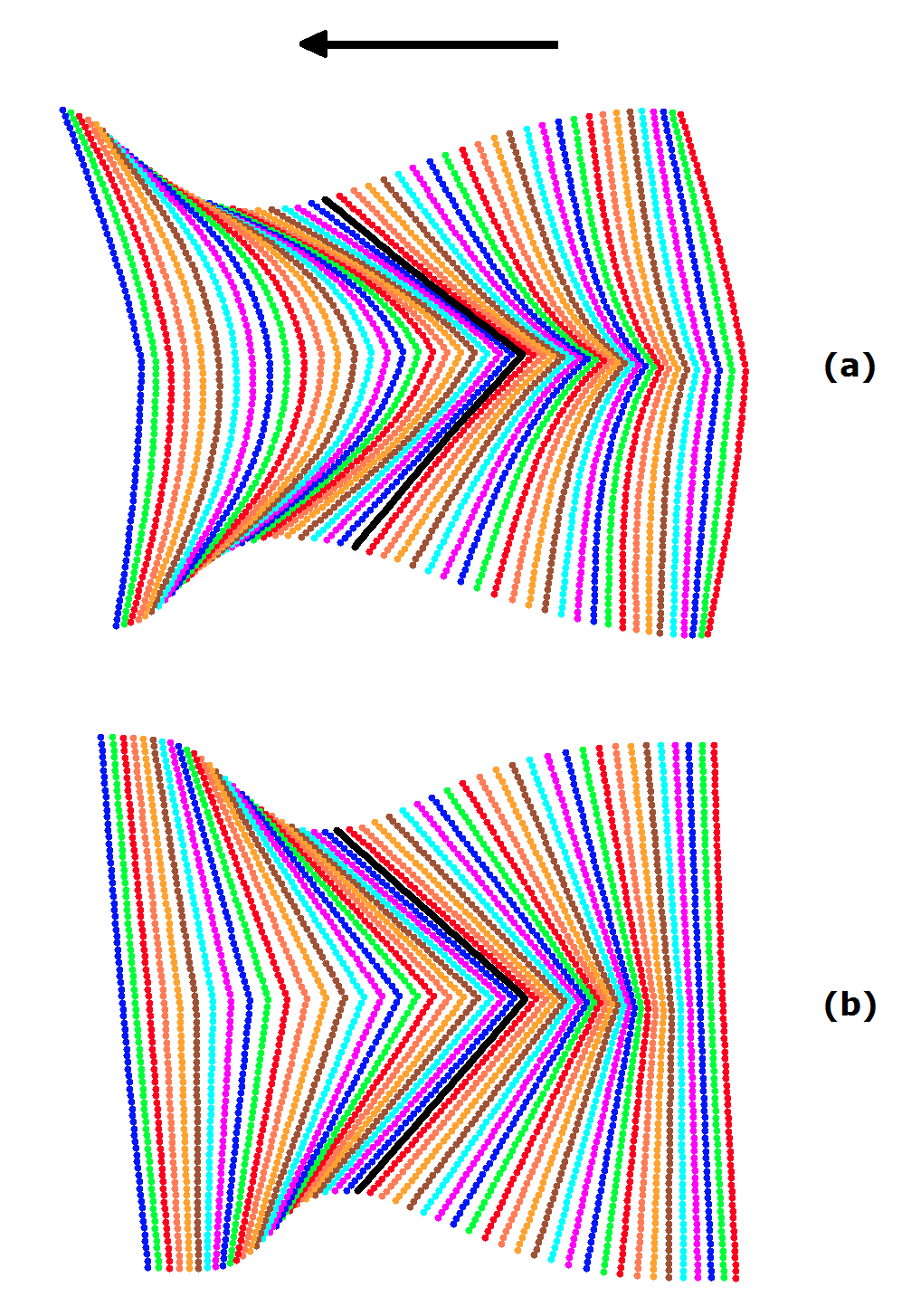}
\caption{Sequence of conformation of the arms (here different colors are used just for the better visualization) of the one-hinged swimmer over one complete beat cycle, (a) represents the case for a flexible arms and (b) for the case of rigid arms. From right to left, we observe the sequence of closing, from the center bold conformation shows the opening of the arms of the swimmer. The black arrow at the top shows the direction of swimming for the flexible swimmer.}
\label{fig2}
\end{figure}

	From the Scallop theorem, we know that the movement undergoing time reversal symmetry will not be able to propel in a low Reynolds number fluid. In the present work we simulate a two-dimensional one-hinge swimmer similar to a two-dimensional scallop. Figure \ref{fig2}(a) shows the shape conformation when we have flexible arms and figure \ref{fig2}(b) is when we have the rigid arms, for one complete cycle. During one complete cycle the curvature of the hinge undergoes a change as $\sin^2 {(2\pi\nu t)}$, which means we have only positive cycles. In figure \ref{fig2}(a) when we look from right to left, first half of the figure,  (till the solid black line), $\sin^2 {(2\pi\nu t)}$ changes from $0-1$, shows the closing of the arms and the second half shows the opening of the arms with $\sin^2 {(2\pi\nu t)}$ changing from $1-0$. The curvature of the hinge changes as a function of the time ($\alpha = A\sin^2 {(2\pi\nu t)}$) as given in the equation \ref{eq.5}. When $t=0$ we have zero curvature that means arms of the swimmer is opened to the maximum possible extent or we start from a straight line. As time progresses, $\alpha$ increases and the arms of the swimmer start to close from the center of the swimmer, similar to a two-dimensional scallop. When the arms start to close, the center of mass of the swimmer moves in the backward direction. The rest of the arms follow the actuation of the hinge points. As the actuation only happens on the hinge, there will be a delay for the actuation to reach the end of the arms. This will cause the points close to the hinge of the swimmer to close faster compare to the points further away. So as the arms close, the ends bend in an outward direction and resembles ${\bf V}$ shaped conformation. When it reaches $\sin^2 {(2\pi\nu t)}=1$, we get a maximum value of curvature equal to the amplitude $A$. After that the curvature at the hinge starts decreasing and the arms of the swimmer start to open up. Again the point away from the hinge will be moving in the closing direction, due to the delay in the propagation of the actuation. By that time the points close to the hinge would have already started to move in the opening direction. Due to this the ends of the swimmer are now bend inward as can be observed in the latter half of the figure \ref{fig2}(a). The center of mass of the swimmer also starts moving in the forward direction and the ends of the arms start to retract now in a ${\bf U}$ shaped conformation. At a later time, when the arms reach $\sin^2 {(2\pi\nu t)}=0$ or $\alpha=0$, the arms of the swimmer are opened at maximum possible extent. As the swimmer has a $V$ shape conformation during closing of the arms while in case of opening it has a $U$ shaped conformation, thus breaking the time inversion symmetry and the one-hinge swimmer move ballistically from right to left as shown by the black arrow in figure \ref{fig2}, which was also predicted by E. Lauga \cite{lauga2007floppy}. The propulsion of the one-hinged swimmer is illustrated in the supplementary materials movie1. 
When we keep the bending stiffness $K_b=10^7$, in our simulation we obtain rigid arms for one-hinge swimmer. In the case of rigid arms, the oscillations at the hinge are followed by every point of the arms and thus the swimmer performs reciprocal motion. There is no difference in the opening and closing conformation, as can be observed in figure \ref{fig2}(b). As a consequence the center of the mass of the swimmer does not undergo ballistic motion, as illustrated in the supplementary material movie2. 

\subsection{Mean squared displacement}
\subsubsection{Finite size effect}
\begin{figure}
\centering
  \includegraphics[scale=0.23]{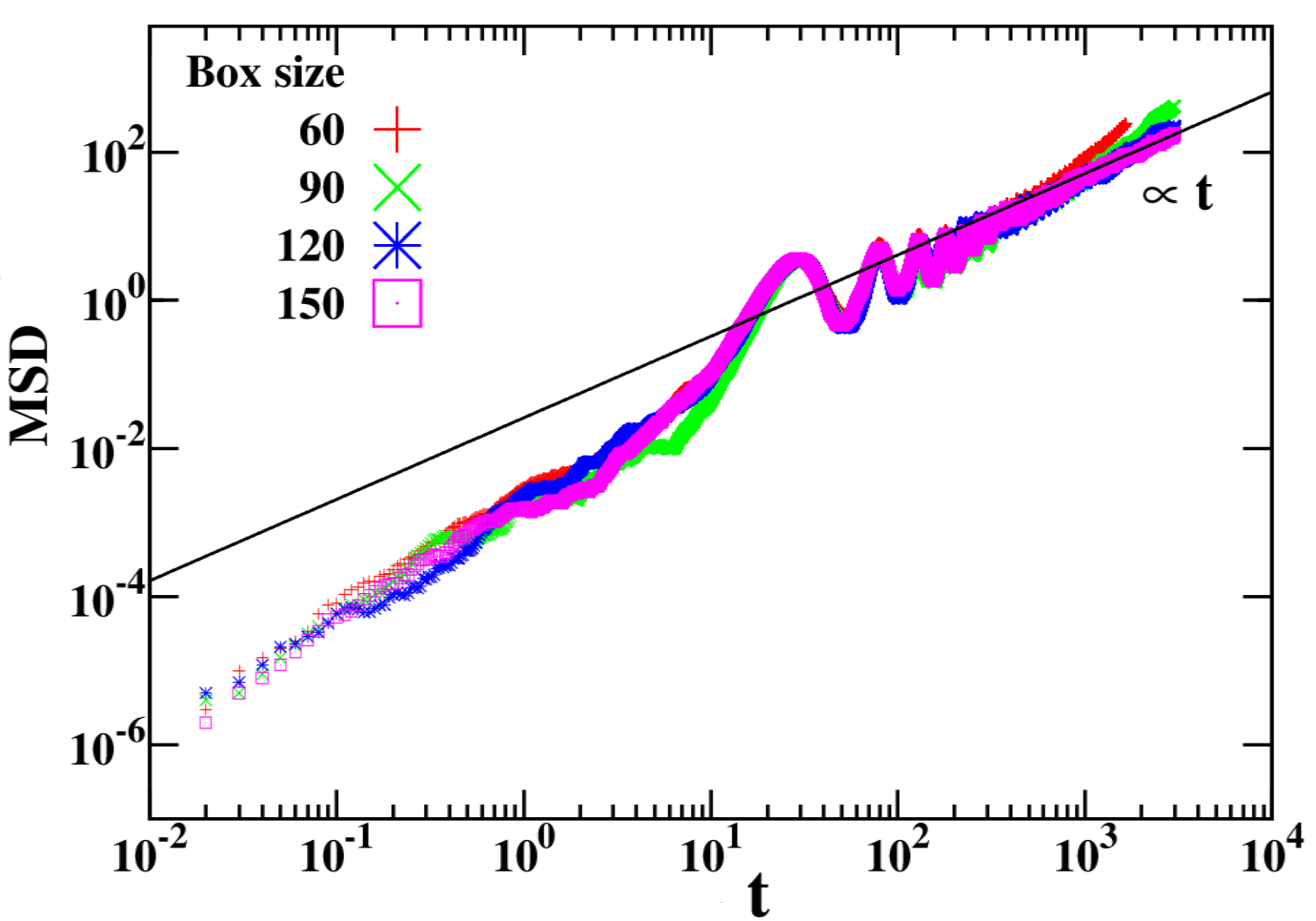}
\caption{MSD of a rigid scallop of arm length $L=15$ in four different boxes (box sizes are mentioned in figure). Here solid black line has slope one.}
\label{fig01}
\end{figure}

In problems involving hydrodynamics effect the simulation box has to be choosen appropriately to avoid finite size effect \cite{PhysRevE.85.021901,Sperm_mpc_2008}. To verify the effect of finite box size we have calculated the mean square displacement of the center of mass (MSD) of the rigid arm swimmer  
\begin{equation}
\langle {\bf R}^2 \rangle= \langle {\left({{\bf r}_{cm}(t)-{\bf r}_{cm}(t_0)}\right)^2} \rangle.
\end{equation}
Where ${\bf r}_{cm}(t_0)$ and ${\bf r}_{cm}(t)$ are the position of the center of mass of the swimmer at beginning of the simulation $t=0$ and at time $t$ respectively. In figure \ref{fig01} we have plotted the MSD as a function of time $t$ for a swimmer with length of one arm $L=15$ for $4$ different box sizes. The rigid arms scallop will only diffuse and we expect slope of $1$ of MSD, due to the inherent thermal fluctuation of the MPC fluid. In figure \ref{fig01}, we can see that in all the cases, till $t\approx 10$ swimmer does hardly move, because bending waves take some time to propagate from hinge to the end of the arms. As we have discussed in the previous section, initially a swimmer has open arms and with time it starts to close the arms due to that center of the mass displaces, we get an upswing in MSD. Once the curvature reaches the maximum, arms starts to open again and center of the mass tries to come back to the previous position and due to that we get a down swing in MSD. Because of the oscillatory stroke that the swimmer performs, MSD of the swimmer always oscillates. We observe that when the box size is $60$ or $4$ times the length of the arm $L$, after $1-2$ cycles of oscillation MSD attains slope of $1$. When the MSD reaches a time of $t>300$, we observe a deviation from the slope of $1$. When the size of the box was increased to $90$ or $6$ times of $L$, we observe that the deviation of MSD from slope of $1$ happens at a later time $t>10^3$, indicating that the finite size effect happens at a much later time. When we further increased the size to $8$ and $10$ times the length of the arm of the swimmer we did not observe a deviation from a slope of $1$ upto the time of our observation. So in the present work we have always kept the size of the simulation box $10$ times the length of the arm of the swimmer to avoid finite size effect.

\graphicspath { {Result/1/} }

\subsubsection{Ballistic motion}
\begin{figure}
\centering
  \includegraphics[scale=0.175]{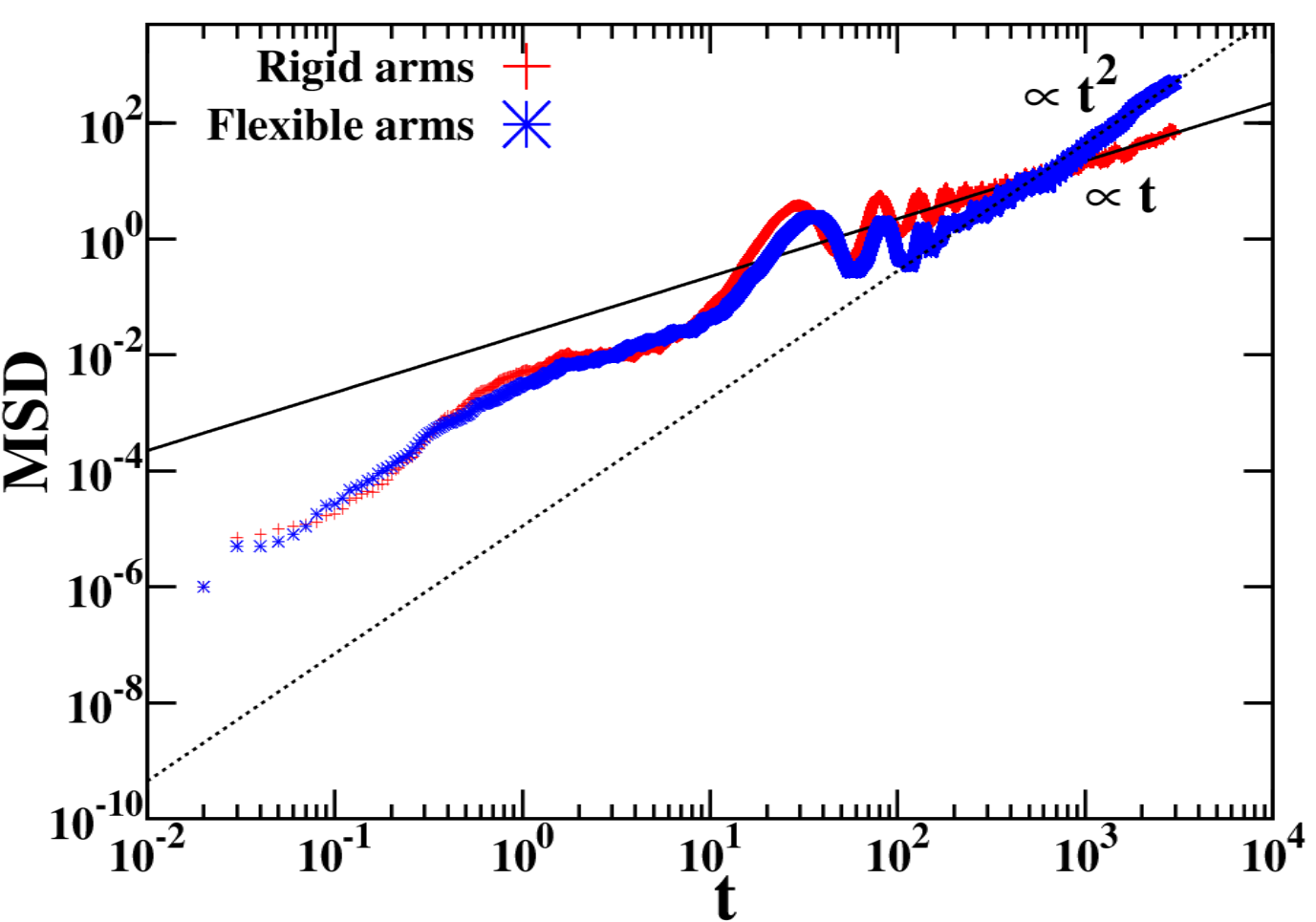}
\caption{Mean squared displacement of the center of mass position of the swimmer is plotted as a function of time $t$ for the flexible arms (blue) and rigid arms (red). The solid line has slope of $1$ indicating diffusive motion and dotted line has slope of $2$ representing directed motion of the swimmer.}
\label{fig3}
\end{figure}

In order to obsereve the difference between a flexible and a rigid scallop we have calculated the MSD as a function of time for both the case as shown in figure \ref{fig3}. Here we observe that after $3-4$ cycles of oscillation the flexible swimmer starts to undergo ballistic motion characterized by the slope of $2$, with the swimmer moving toward the left as indicated by the arrow in the figure \ref{fig2}(a), as predicted by E. Lauga \cite{lauga2007floppy}. As mentioned before for the rigid arms the center of mass of the swimmer undergoes diffusive motion characterized by $<\textbf R^2> \enspace \propto \textbf t$.
\graphicspath {}
\subsection{Effect of Flexibility}
\begin{figure}
\centering
  \includegraphics[scale=0.25]{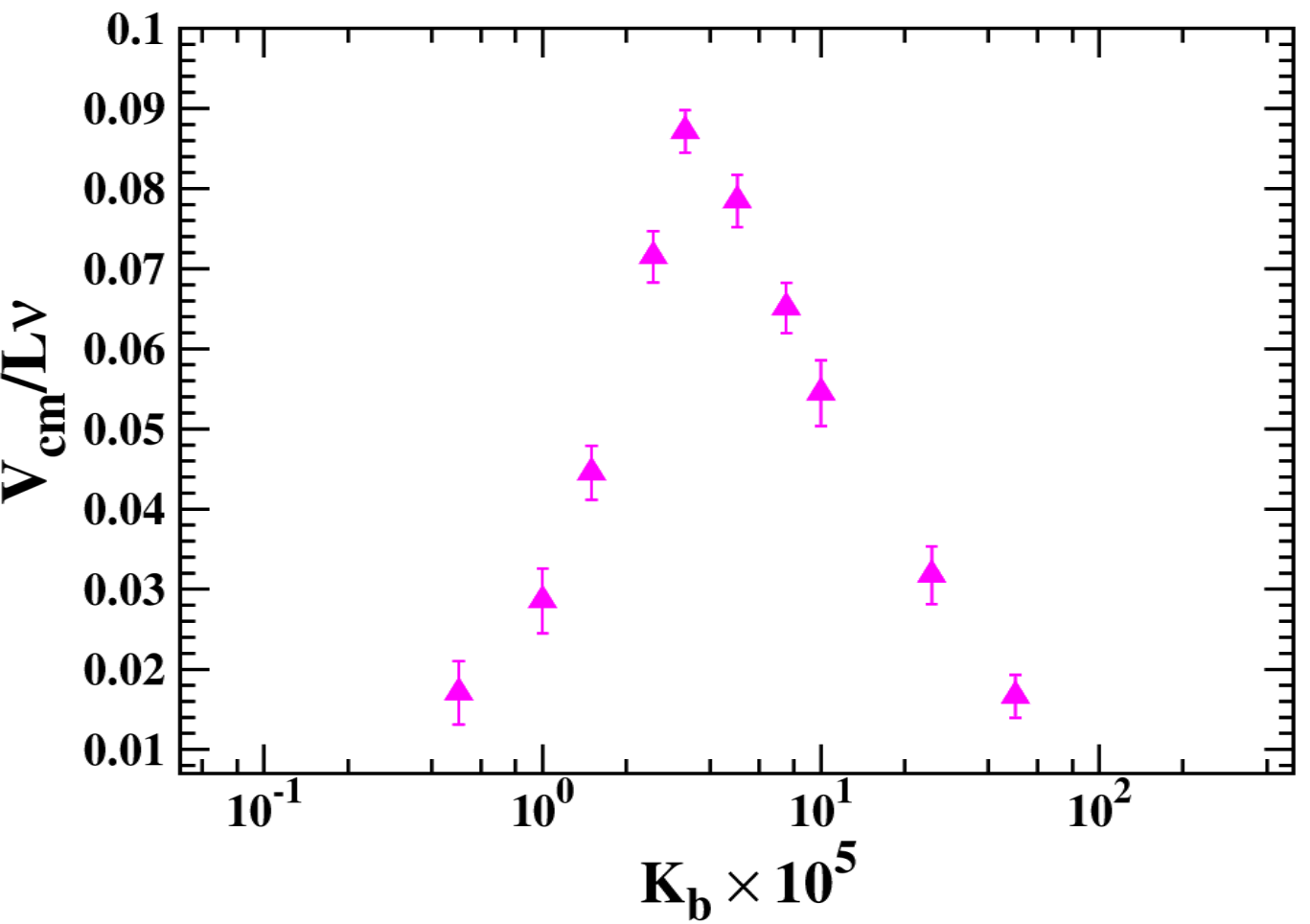}
\caption{The reduced velocity of the swimmer is plotted as a function of bending rigidity of the arms of the swimmer.}
\label{fig4}
\end{figure}

From the previous section we know in the present work flexibility plays a major role in the swimming of the two arms swimmer. To study the effect of flexibility we measure the reduced velocity after the swimmer has attained ballistic motion which is defined as $V_{cm}/L \nu$, where $V_{cm}=\sqrt{\langle R^2 \rangle}/t$. The Reynolds number is defined $Re = V_{max}L/\mu$  \cite{Purcell_1977,Sperm_mpc_2008,padding2006hydrodynamic}, we use $ \mu=3.6$ and the maximum velocity we have studied for the scallop is $V_{max} \approx 0.03$, the length of the arm of the scallop is $L=20$, so the Reynolds number of our system $Re<0.16$, which is consistent with the previous studies of low Reynolds number regime using MPC simulations \cite{Sperm_mpc_2008,AndreasPhysRevLett,munch2016taylor}. The Peclet number is defined as $Pe = V_{cm}L/D$, where $D$ is the diffusion coefficient of the rigid swimmer \cite{padding2006hydrodynamic} and in  our case the range of $Pe = 16–80$, which means the thermal fluctuations do not play a significant role in our simulations \cite{Elgeti}. In MPC simulation the Mach number of the fluid should be very small in order for the fluid to be incompressible. The Mach number is defined as $Ma = V_{max}/V_{sound}$, here $V_{sound}= \sqrt{2}$ is the speed of sound in $2$ dimension for MPC fluid. In our case for the  maximum frequency we have used the $V_{max} \approx 0.03$ and so $Ma \approx 0.021$ \cite{PhysRevE.85.021901}, which is within the incompressible limit of MPC.  In order to verify that MPC fluid is in the incompressible limit at all length scales in our simulation we calculated the mach number for the tip of the swimmer which moves with the maximum velocity $\approx 0.24$. The mach number for the tip of the arm turns out to be $Ma \approx 0.17$, which is again within the accepted MPC limit for having an incompressible fluid \cite{padding2006hydrodynamic,re0231fId0,refId0}. In figure \ref{fig4}, we plot the reduced velocity as a function of bending rigidity ${K_b}$ of the arms of the swimmer. When $K_b=0$ and $K_w=0$ our swimmer becomes passive and behaves as a Gaussian polymer chain. When we increase ${K_b}$ from $10^{5}$ onward, keeping $K_w=4 \times 10^5$, we observe a nominal increment in speed. The reason being that here the arms of the swimmer are very flexible and the wave which passes from the hinge is damped before reaching the end point. Here, the configurations of the swimmer do not change appreciably during the entire beat cycle. The arms of the swimmer become progressively rigid by increasing $K_b$, the speed increases as shown in figure \ref{fig4}. Here we have a competition between the elastic forces of the swimmer and viscous drag because of the fluid particles. The speed starts to increase till it reaches a maximum when the elastic forces cancel the frictional forces, for intermediate bending stiffness $K_b= 3.25 \times 10^5$. On further increasing the rigidity of the arms, the velocity goes down. Near $K_b=10^7$ it starts behaving as a conventional scallop and eventually the speed of the scallop goes towards zero as we keep on increasing the rigidity of the arms.

\graphicspath {}
%\subparagraph{}
 
\subsection{Beating frequency}
\graphicspath { {Result/1/} }
\begin{figure}
\centering
  \includegraphics[scale=0.178]{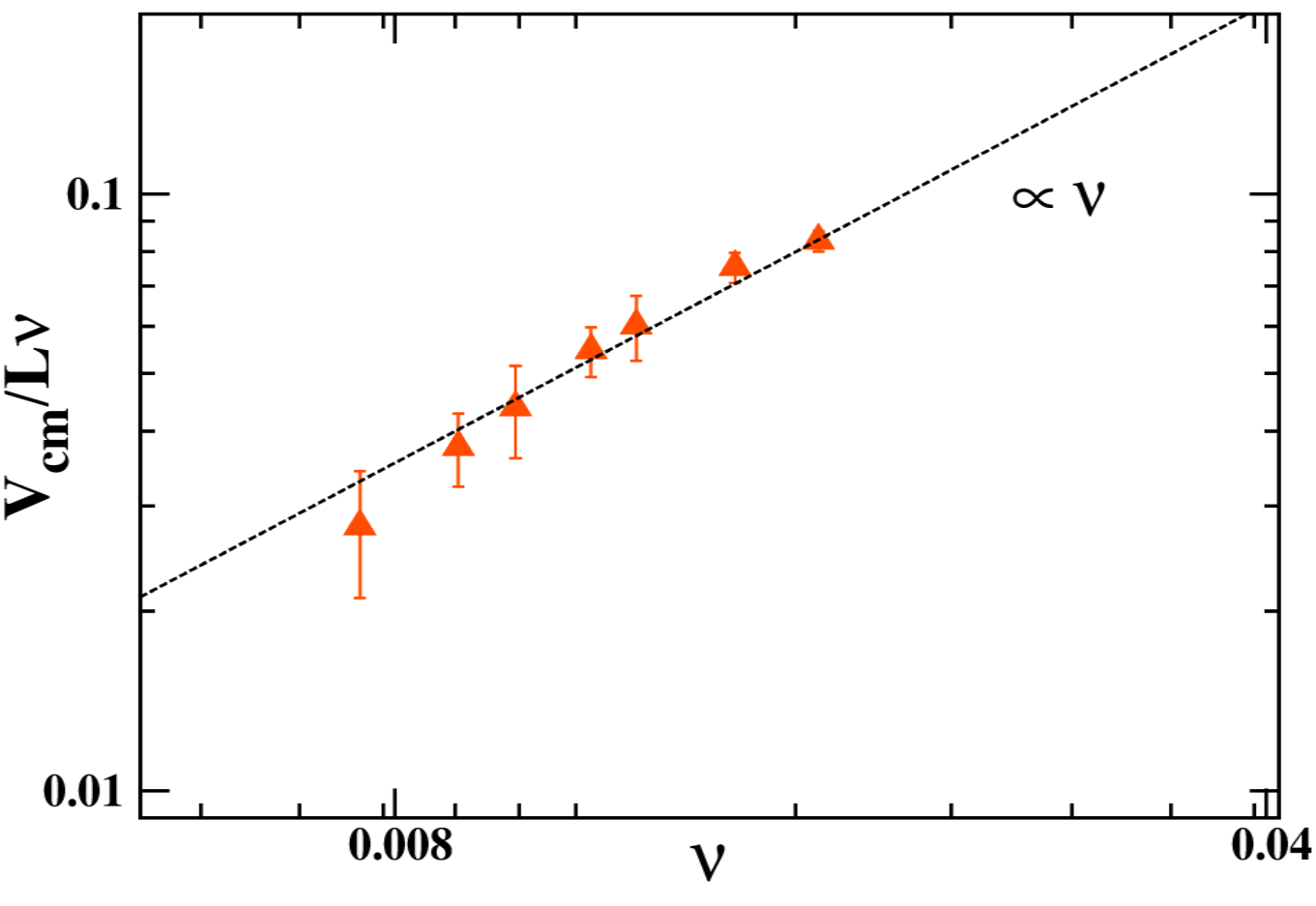}
\caption{The reduced velocity is plotted as a function of frequency of the actuation. The solid line has slope of one as predicted for the small amplitude approximation.}
\label{fig5}
\end{figure} 
In figure \ref{fig5}, we have plotted the reduced velocity of the swimmer with respect to the beating frequency $\nu$ imparted on the hinge. In the present study for the one hinge swimmer, we are considering only large amplitudes, the reduced velocity scales linearly with $\nu$, as shown by the straight dashed line in the figure \ref{fig5} which has slope of $1$. For a single-armed swimmer Wiggins et al. had derived the equation for velocity using small amplitude approximation, where it was shown that the reduced velocity scales linearly with $\nu$. In our case, the amplitude we have considered $A=4.25$ is large, still we observe that the scaling is in agreement with that of Wiggins et al. \cite{Goldstien_1998}.  

\subsection{Amplitude of the actuation}
\begin{figure}
\centering
  \includegraphics[scale=0.175]{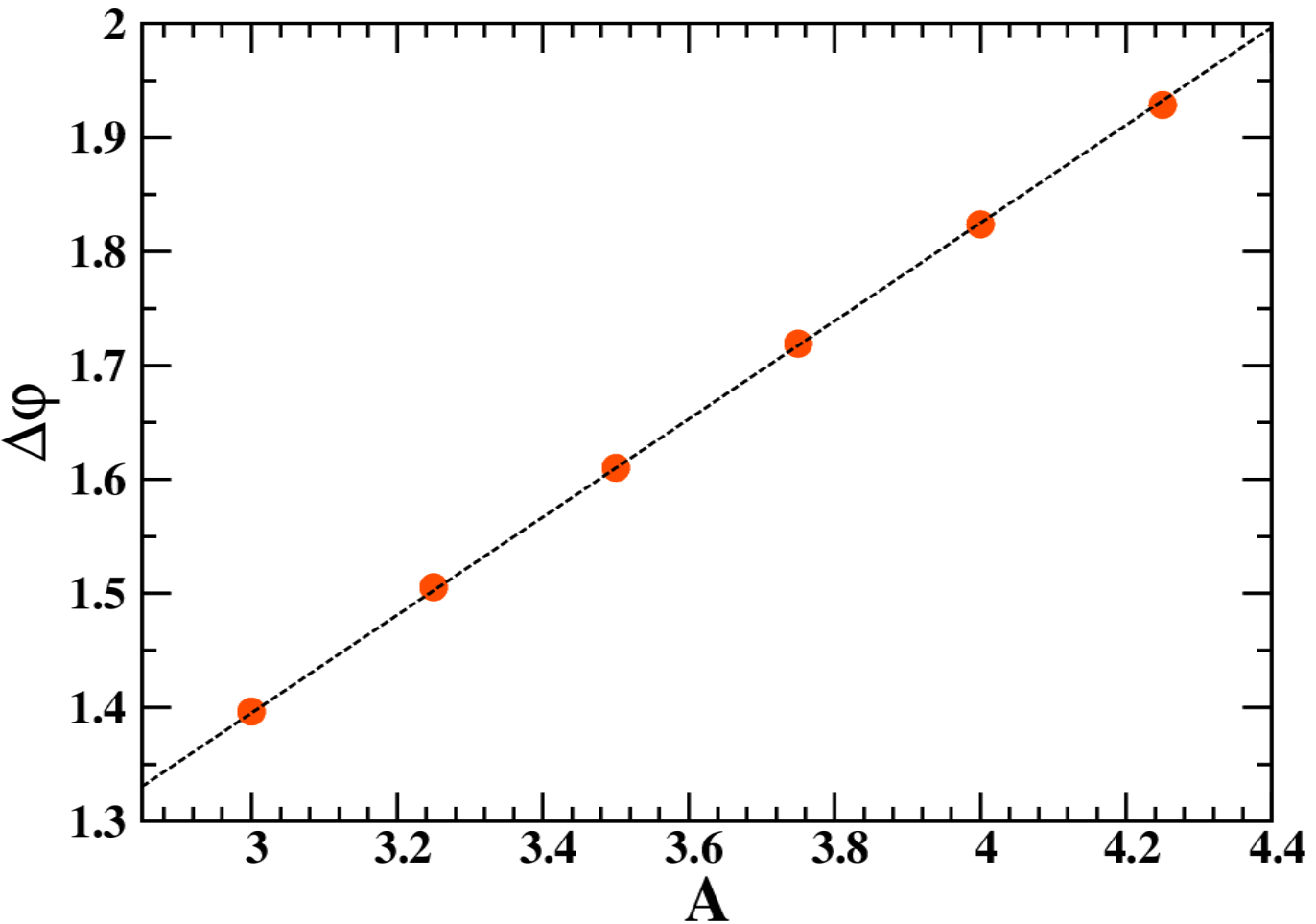}
\caption{$\Delta\varphi$ (in radian) is plotted as a function of the amplitude of actuation. The dashed line shows the linear fit given by $(0.43 A/a_0 + 0.1)$.}
\label{fig6}
\end{figure}

For closing and opening of the arms of the swimmer, the parameter that we are changing is the amplitude $A$ of the bending wave given in the equation \ref{eq.5}. The amplitude is chosen in such a way whereby we make sure that the mass points of the swimmer close to the hinge will never overlap. As we change the amplitude of the beating, the angle created between the arms $\varphi$ of the swimmer changes as well, see figure \ref{fig1}. The maximum angle between the arms in the simulation nearly stays around $\varphi_{max}= \pi$, while the minimum angle varies between $\varphi_{min}=\pi/3 - 2\pi/3$. In the present study the amplitude is related to the difference in the minimum and maximum angle made by the arms of the swimmer. In figure \ref{fig6}, we plot the amplitude $A$ as a function of $\Delta \varphi= \varphi_{max}-\varphi_{min}$, which is the difference between the minimum and maximum angle during the opening and closing of the arms of the swimmer.

 When $\Delta \varphi = \pi$, then the maximum angle will be $\pi$ and the minimum angle will be $0$ between the arms of the swimmer, which means the arms of the swimmer is going to overlap at the minimum angle. For measuring $\varphi$, we calculate a unit vector between the hinge and the fourth bead from the hinge for the left arm $\hat{p}_l$ and the right arms $\hat{p}_r$ respectively. Then the angle $\varphi=\cos^{-1} ({\hat{p}_l  \cdot \hat{p}_r})$, also note that for the beating of the arms we are changing the spontaneous curvature of the beads close to the hinge by $\sin^2{(2\pi \nu t)}$, which means the angle between the arms will not go beyond $\pi$. The dashed line in figure \ref{fig6} is drawn from straight line equation which shows that $A \propto \Delta \varphi$. In the present work we have kept $K_w\approx 10^5$, the we observe that arms can open and close in such a way $\Delta \varphi> \pi/3$. 
\subparagraph{}

\begin{figure}
\centering
  \includegraphics[scale=0.1825]{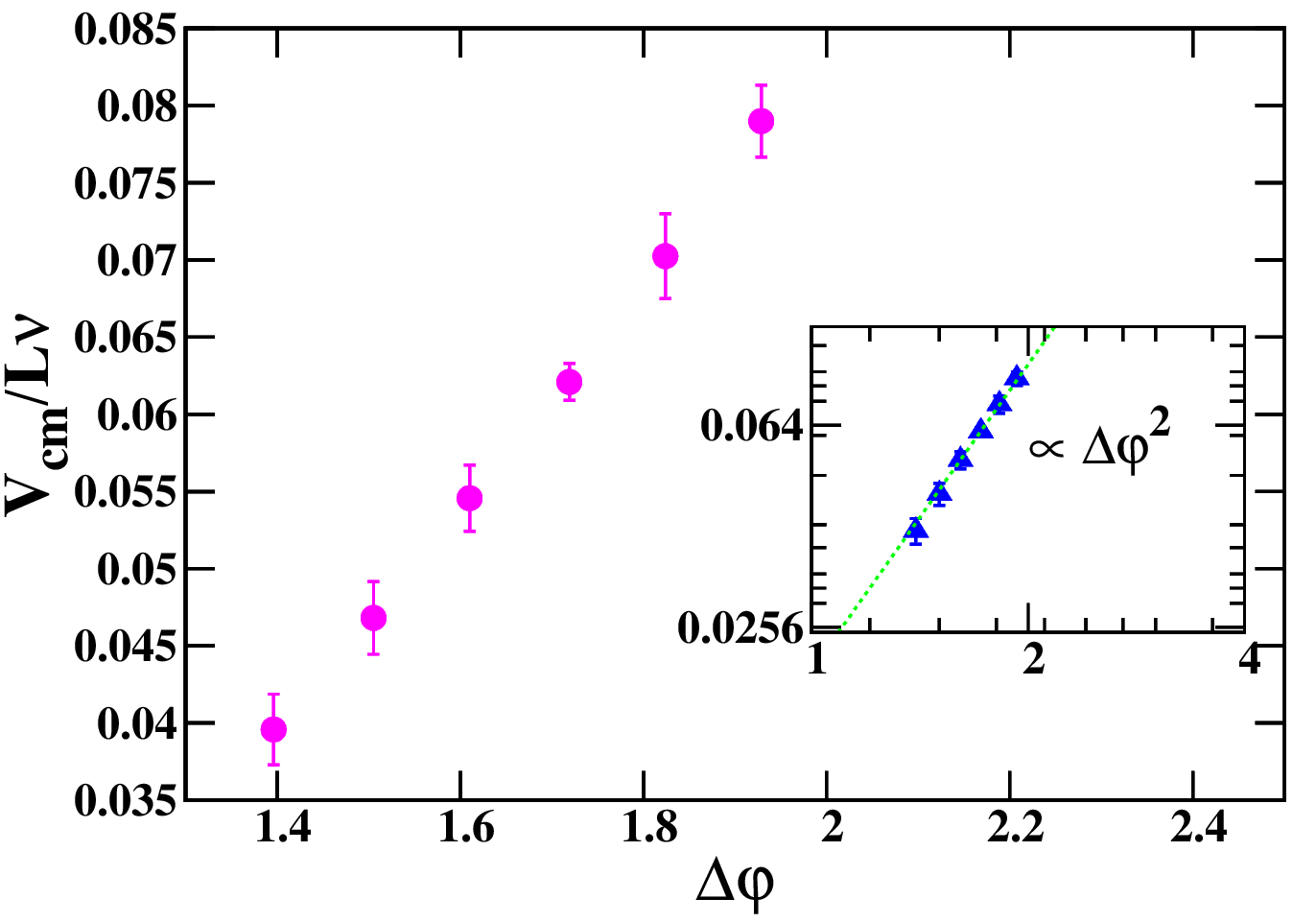}
\caption{Reduced velocity of the swimmer is plotted as a function of $\Delta\varphi$ (in radian), the dashed line (inset) has slope of two as predicted for small amplitude approximation.}
\label{fig7}
\end{figure}

In figure \ref{fig7}, we plot the reduced velocity as a function of the $\Delta \varphi$. Here we observe that as the difference in the angle between the arms increases the speed of the swimmer also increases. We observe that the reduced velocity scales as the square of $\Delta \varphi$, the dashed line in the inset of figure \ref{fig7} has slope of two. The equation for the velocity has been deduced from the small amplitude approximation for the single-armed swimmer \cite{Goldstien_1998} as well as for the one-hinge swimmer \cite{lauga2007floppy}. In these works they had shown that the reduced velocity scales with the square of the amplitude of the actuation. Even though our amplitudes are larger than that were considered in these work, we observe that the reduced velocities have similar kind of scaling as that of small amplitude approximation. We were not able to further increase the amplitude of the actuation as that will lead to the overlap of the arms of the swimmer, which is not a physical scenario for an artificial swimmer.
\graphicspath {}

 \graphicspath {}

\subsection{Sperm number}
\graphicspath {}
\begin{figure}
\centering
  \includegraphics[scale=0.185]{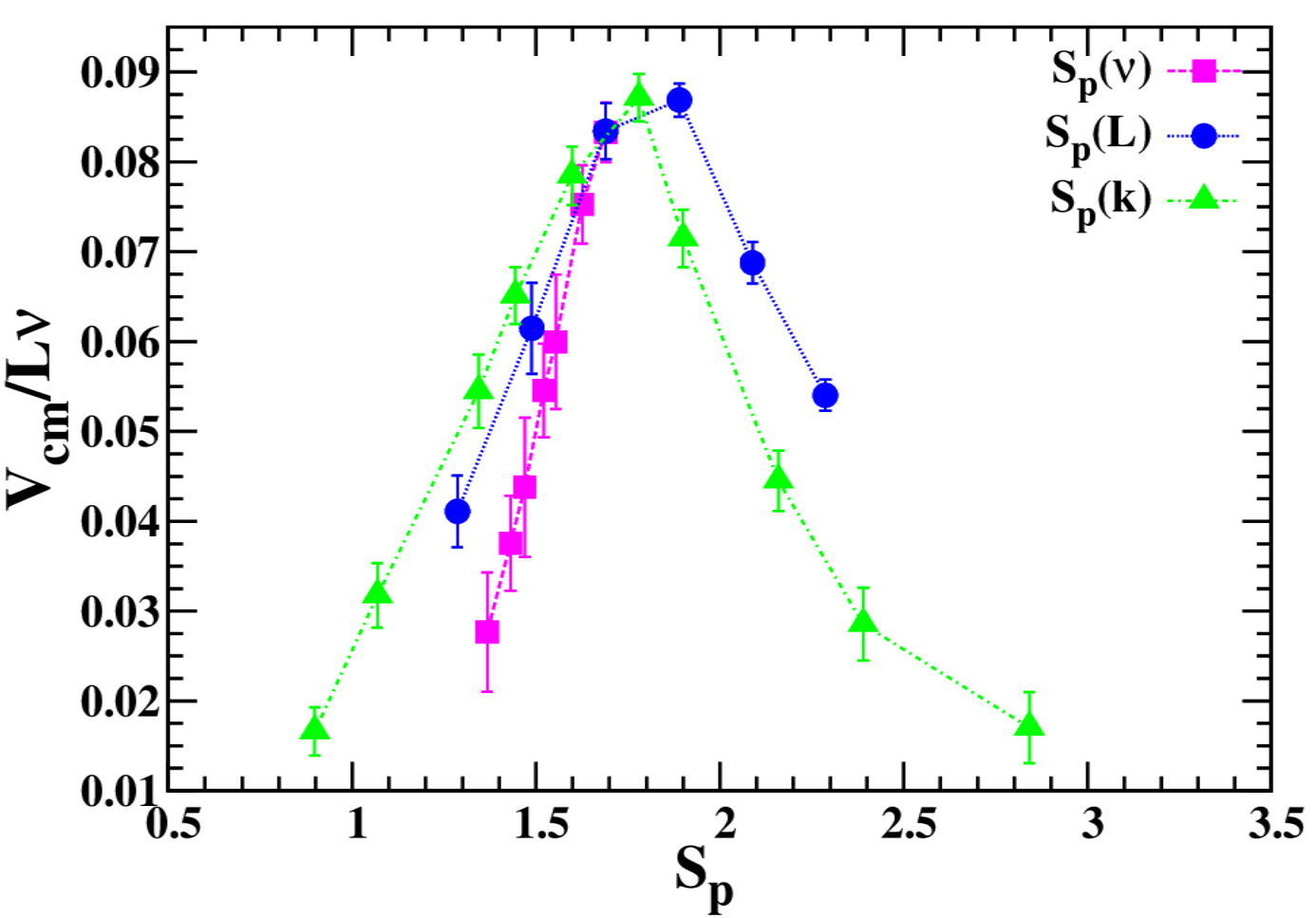}
\caption{The reduced velocity of the swimmer is plotted as a function of Sperm number. The value of $S_p$ was varied, changing $\nu$ (square), $L$ (circle) and $k$ (triangle) keeping all other parameters the same.}
\label{fig8}
\end{figure}
Wiggins et al. had shown that if a flexible rod is perturbed at one end by $y_0 \cos \omega t$, for small amplitudes, the dynamics of the one arm follows the dimensionless hyper diffusion equation $\frac{dy}{dt}=S_p^4 \frac{d^4y}{dx^4}$, where $S_p$ is a dimensionless number called Sperm number. The Sperm number is given by 
\begin{equation} 
S_p =  L \left( \frac{\xi_\perp 2 \pi  \nu}{k} \right)^{\frac{1}{4}}. 
\label{e.7}
\end{equation}
 Where $L$ is the length of the arms of the swimmer, $k=K_b l_0$ is stiffness of the arms of the swimmer, $\nu$ is the frequency and $\xi_\perp$ is the frictional coefficient per unit length in perpendicular direction of motion defined to be consistent with the one-armed swimmer. In the present study, we calculate the perpendicular friction coefficient using $\xi_\perp=4\pi\eta / [\ln{(L/r)}+1/2]$  \cite{pak2015theoretical}, where $\eta$ is the viscosity of the fluid and $r$ is the radius of the arm. In the present work, the spring that connects two beads behave as a rigid rod of length $l_0=0.5 $, as a result we have considered the radius of the rod to be $r=0.25$.  In figure \ref{fig8} we show the reduced velocity of the swimmer as a function of the Sperm number. For the small value of the $S_p$ we observe that the elastic forces dominate and we get a small velocity, while in the high Sperm number region the drag forces acting on the arms of the swimmer dominate and again the velocity goes down. As there is a competition between the viscous and elastic forces in the arms of the swimmer, there should be one particular $S_p$ value where we should observe a maximum for the velocity, which is around $ \sim 1.8$ in our case. For the one-armed swimmer with small amplitude approximation, it was shown that the maximum velocity is when $S_p\approx 4$, while Lagomarsino et al. have demonstrated using simulations that for the same one-armed swimmer with large amplitude, the maximum velocity happens when $S_p\approx 2$ \cite{lagomarsino2003simulation,lowe2003dynamics}. This is similar to our present work, where we consider the amplitude larger than the small amplitude approximation. In figure \ref{fig8}, $S_p$ is varied by changing the length of the arms, the frequency of actuation and the stiffness $K_b$ of the arms of the swimmer keeping the rest of the parameter constant of the equation \ref{e.7}. Here we can see that the peak positions of the reduced velocities remain almost same. For the small values of the Sperm number till $S_p< 1.6$, we find continuous increment in scaled velocity with $S_p$ as expected. After the maximum, we observe that the scaled velocity continuously decreases as we increases the $S_p$ and after $S_p>2$ we observe a slow drop in scaled velocity. For small amplitude approximation Wiggins et al. had shown that the scaled velocity attains a plateau for higher $S_p$ value. Our results are similar to the results obtained by Lagomarsino et al. \cite{lagomarsino2003simulation}, where they showed that for large values of $S_p$, when large amplitudes are considered, the velocity does not stay constant, it slowly decreases with $S_p$. 

\section{DISCUSSIONS} 
\label{sec:4}
It was shown that the one-armed flexible swimmer \cite{magnetic_swimmeri_nature_2005,Gauger_2006} will undergo ballistic motion only if a passive head is attached to the arm. The two-armed swimmer with a head was initially considered by Lauga \cite{lauga2007floppy}, for simplicity of the calculation he ignored the hydrodynamic interactions between the arms of the swimmer and also the velocity of the swimmer was derived under the infinitely small amplitude approximation. Lauga has also shown analytically that if two arms are actuated similar to a one-armed swimmer a passive head is not required to break the time inversion symmetry. In the present work we are able to show that if two filaments are joined by a hinge and actuated only at the hinge the artificial swimmer starts to behave as a self propelled object. We obtain the direction of the swimmer as predicted by Lauga \cite{lauga2007floppy} but the magnitude of velocity is not in agreement with our work. This may be because in the present work there are hydrodynamic interactions between the arms of the swimmer as well as the amplitude of the actuation in the present work is higher than that considered for the small amplitude approximation.  Also recently Tian Quie et al. \cite{qiu2014swimming} demonstrated that a micro scallop can swim in a non Newtonian fluid. In this work they have used a rigid arm scallop in a non Newtonian fluid while in the present work our scallop or one-hinge swimmer have flexible arms thereby performing the ballistic motion even in Newtonian fluid. It is known that swimmer African Trypanosome velocity is increased approximately $8$ times when it swims in a fluid (non Newtonian) having obstacles of size and spacing of RBC in blood \cite{heddergott2012trypanosome} and it will be interesting to study the swimming of the one hinge swimmer in these environments.
   
   Recently Haug et al. have shown that they can create soft robotic materials using light actuated materials \cite{huang2015miniaturized}. The light-driven liquid-crystal (LDLC) material \cite{yu2003photomechanics}, which are sensitive to ultra-violet light, can convert light into the mechanical energy with quick response and large deformation. When we shine UV light the LDLC undergoes a phase transition by which it gets converted into a shorter molecule, while shining visible light it recovers its original conformation. We believe using these materials experimental realization of the one-hinge swimmer is possible. They have already demonstrated using these materials that they are able to recover the shape conformation for a one-armed swimmer and to extend it to a one-hinge swimmer should not be too challenging.
\section{CONCLUSIONS}  
\label{sec:5}
In this paper, we have modeled a two-dimensional scallop or a one-hinge swimmer. We have shown that the hydrodynamic interactions between the mass points of the swimmer and the fluid particles can be simulated using multi-particle collision dynamics (MPC). We have also demonstrated that if the arms of swimmer are very rigid, it follows the Scallop theorem and the swimmer is not able to propel itself through the viscous fluid. While when the arms are made flexible the time inversion symmetry is broken and swimmer performs ballistic motion. We have also shown that the velocity of the swimmer has a maximum, for intermediate bending rigidity along the arms of the swimmer. For small bending rigidity the arms are very flexible and the actuation is not able to produce the desired shape conformations, while for stiff arms we are closer to the Scallop theorem and the velocity goes towards zero. 
The reduced velocity is also studied as a function of frequency as well as the amplitude of actuation and we have shown a similar scaling relation as predicted by elastohydrodynamic within the small amplitude approximation even though we have considered large amplitude in the present work. We were also able to define Sperm number for the swimmer and also showed that the reduced velocity had a maximum at $S_p\sim 1.8$ consistent with what is expected for large amplitude actuation.
\subsection*{Acknowledgements}
We would like to thank DST for funding and also would like to acknowledge IIT Delhi HPC facility for computational resources.

\end{document}